# Decentralized Social Media and Artificial Intelligence in Digital Public Health Monitoring

Marcel Salathé[1], Sharada P. Mohanty[1]

[1]Digital Epidemiology Lab, School of Life Sciences, School of Computer and Communication Sciences, EPFL, Switzerland

## Abstract

Digital public health monitoring has long relied on data from major social media platforms. Twitter was once an indispensable resource for tracking disease outbreaks and public sentiment in real time. Researchers used Twitter to monitor everything from influenza spread to vaccine hesitancy, demonstrating that social media data can serve as an early-warning system for emerging health threats. However, recent shifts in the social media landscape have challenged this data-driven paradigm. Platform policy changes, exemplified by Twitter's withdrawal of free data access, now restrict the very data that fueled a decade of digital public health research. At the same time, advances in artificial intelligence, particularly large language models (LLMs), have dramatically expanded our capacity to analyze large-scale textual data across languages and contexts. This presents a paradox: we possess powerful new AI tools to extract insights from social media, but face dwindling access to the data. In this viewpoint, we examine how digital public health monitoring is navigating these countervailing trends. We discuss the rise of decentralized social networks like Mastodon and Bluesky as alternative data sources, weighing their openness and ethical alignment with research against their smaller scale and potential biases. Ultimately, we argue that digital public health surveillance must adapt by embracing new platforms and methodologies, focusing on common diseases and broad signals that remain detectable, while advocating for policies that preserve researchers' access to public data in privacy-respective ways.

# Introduction

Social media has fundamentally changed how we measure and understand population health behaviors in real time. Over the past decade, social media platforms like Twitter enabled a new form of health monitoring, mining online information flows to assess public health trends. Twitter in particular became a vast, real-time data repository for digital public health monitoring, wherein millions of users' posts could be analyzed to track disease outbreaks, public concerns, and health behaviors. A 2021 systematic review catalogued 755 studies using digital data for public health surveillance, spanning communicable diseases, vaccine attitudes, chronic conditions, and mental health [1].

Critically, Twitter's openness and scale made it an attractive resource. The platform's public API historically allowed academics to retrieve large volumes of tweets for research, and early studies demonstrated the feasibility of using Twitter to detect and even predict disease outbreaks. During the 2009 H1N1 influenza pandemic, analysis of Twitter posts showed that tweet frequencies about flu correlated closely with official infection rates [2,3]. Paul and Dredze developed topic modeling approaches to identify health-related content in tweets and showed strong correlations with CDC surveillance data [4]. These high correlations underscored that social media could reliably indicate real-world disease incidence.

Beyond infectious disease tracking, Twitter also became a barometer for health behaviors and sentiments, notably vaccination attitudes. Early work demonstrated the feasibility of assessing vaccination sentiments through online social media, opening new opportunities for monitoring concerns [5]. Subsequent research mapped how pro- and anti-vaccine sentiment clustered in the Twitter network and how these sentiments spread through social influence [6]. Studies have even identified signals of adverse drug reactions from patient tweets [7]. More recent observational work has provided causal evidence linking exposure to anti-vaccine content on Twitter with reduced COVID-19 vaccine uptake and measurable increases in cases and deaths [8]. These are just a few examples of a substantial body of peer-reviewed literature that established that social media data can serve public health by enabling real-time monitoring of disease spread, public knowledge, risk perceptions, and health behaviors.

This promising link between social media and public health research has been disrupted by recent changes in platform governance and data access. In late 2022, Twitter's acquisition by Elon Musk led to a dramatic shift in data accessibility for researchers. In early 2023, Twitter announced it would end free API access and introduced prohibitively expensive pricing tiers [9,10]. The enterprise API reportedly costs approximately $42,000 per month, far beyond the budget of most academic projects. The cheapest tier allows only a tiny fraction of the data previously freely available [10]. The impacts on research have been severe. A survey of social media scholars found that 60% of Twitter-based projects were stalled, cancelled, or moved to other platforms following the Twitter API shutdown [11]. Long-term efforts, including open datasets, dashboards, and tools built around Twitter data, have been abruptly terminated as data pipelines were cut off.

Two interconnected drivers underlie this data lockdown. First, social media content has become strategically even more valuable in the age of AI. Large language models (LLMs) require vast datasets for training, and public social media posts are core material. Platform owners recognize that unrestricted access to this data by outsiders could diminish their competitive advantage in AI model development. Twitter has updated its terms to explicitly prohibit using its content for training AI models [12]. Reddit similarly announced API fees in 2023, later striking a deal to license posts to OpenAI [13]. The surge of interest in generative AI has incentivized platforms to close off their data. Second, heightened privacy awareness and regulatory pressures contributed to restricted access. The 2018 Facebook–Cambridge Analytica scandal was an important moment, after which Facebook severely limited researcher access [14]. Users also have become more wary about how their posts might be used. Surveys indicate that 38% of respondents use social media less due to privacy concerns [15]. Laws like the EU's GDPR and Digital Services Act impose strict requirements on data handling. The net effect is a paradox: platforms justify data restrictions both as competitive strategy and as privacy protection, leaving independent public health researchers severely constrained.

Ironically, these clampdowns on data access come at the very moment when our ability to extract insights from large text datasets has never been greater. The advent of large language models such as the GPT, Gemini, or Claude family of models, has revolutionized natural language processing. Modern LLMs can perform a wide array of language tasks (summarization, translation, sentiment analysis, classification) often in a zero-shot or few-shot manner, without requiring task-specific training [16,17]. This capability is a major asset for public health analysis. In the past, classifying tweets by vaccination sentiment, for example, required developing labeled datasets and training specialized, often language-specific classifiers. Now, large general models can be prompted to perform these classifications with minimal additional training data. They handle multilingual content far better than earlier approaches, allowing a public health team to analyze social media posts in multiple languages with the same model.

Given the new barriers on mainstream platforms, attention is turning to decentralized social networks as potential lifelines for digital public health monitoring. Two platforms deserve particular attention: Mastodon and Bluesky. Mastodon is a federated, open-source social network that is part of the broader "Fediverse." Instead of one central service, Mastodon consists of thousands of independent servers ("instances") that communicate with each other using a common protocol. It provides a free, open API that allows researchers to retrieve data from individual servers, subject to instance policies and user consent settings [19]. This federated structure means there is no single corporate gatekeeper that can shut off data access globally. Importantly, moving research practices from Twitter to Mastodon is technically straightforward, as the API and data formats are similar [19]. However, Mastodon's decentralized nature demands that researchers engage with community norms, which might be different from one server to the next. For example, some servers explicitly forbid data mining without consent [19,20].

Bluesky launched in 2023 as an independent microblogging network built on the open AT Protocol [21]. Bluesky has embraced researcher access from the outset, with its developers emphasizing open data access. A recent analysis captured interactions of approximately 5 million users over Bluesky's first year, finding network structure and usage patterns strikingly similar to Twitter's [21]. Yet Bluesky differs in important ways: users predominantly share center-left, reliable news sources with little misinformation, though clear opinion clusters have emerged on polarizing issues [21]. These emerging platforms are a glimmer of hope for digital public health monitoring. Researchers can potentially move to them to continue certain types of digital health monitoring, such as tracking flu season signals, assessing vaccine sentiment, or observing health discussions. There is also an ethical appeal: users on these platforms are often more aware of data use implications, and data originates in a decentralized fashion that could reduce concerns about single-company exploitation.

To illustrate the feasibility of this approach, we conducted a pilot analysis of Mastodon data collected between September 2023 and October 2024. We processed posts using the GPT-4o mini model of OpenAI with a structured extraction schema designed to capture sentiment, emotional tone, topical content, and health-related signals. The schema required the model to assess each post across multiple dimensions: overall sentiment (positive, neutral, negative), emotional tones using Plutchik's categories, primary topic, and whether the post contained health-related content, including whether it described personal health experiences or discussed public health policy. Figure 1 presents key findings from this analysis. The health signal density panel reveals that a detectable fraction of Mastodon posts are health-related, with variation over time in whether these reflect personal experiences versus policy discussions. The sentiment and topic panels demonstrate that standard digital epidemiology signals, such as shifts in public mood and attention, remain observable on this smaller platform. Importantly, the volume panel confirms that even with a fraction of Twitter's user base, Mastodon generates sufficient data for continuous monitoring. These preliminary results suggest that the combination of open platform APIs and modern LLM analysis can sustain meaningful public health surveillance, though further validation against ground-truth epidemiological data is needed.

(A)

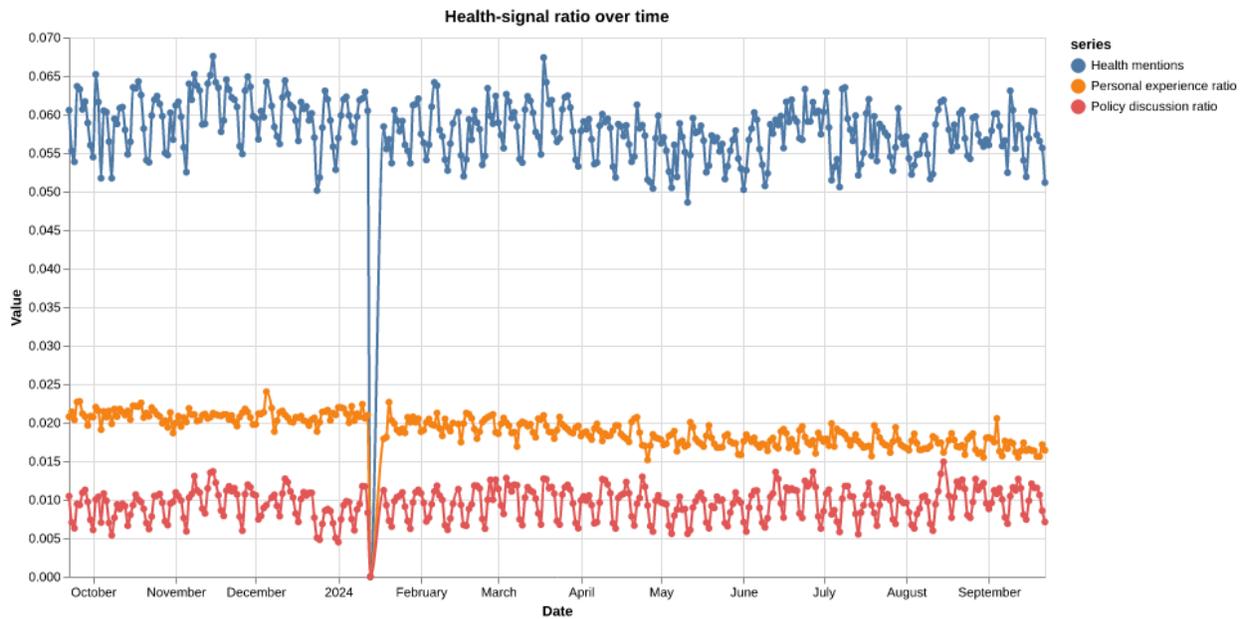

(B)

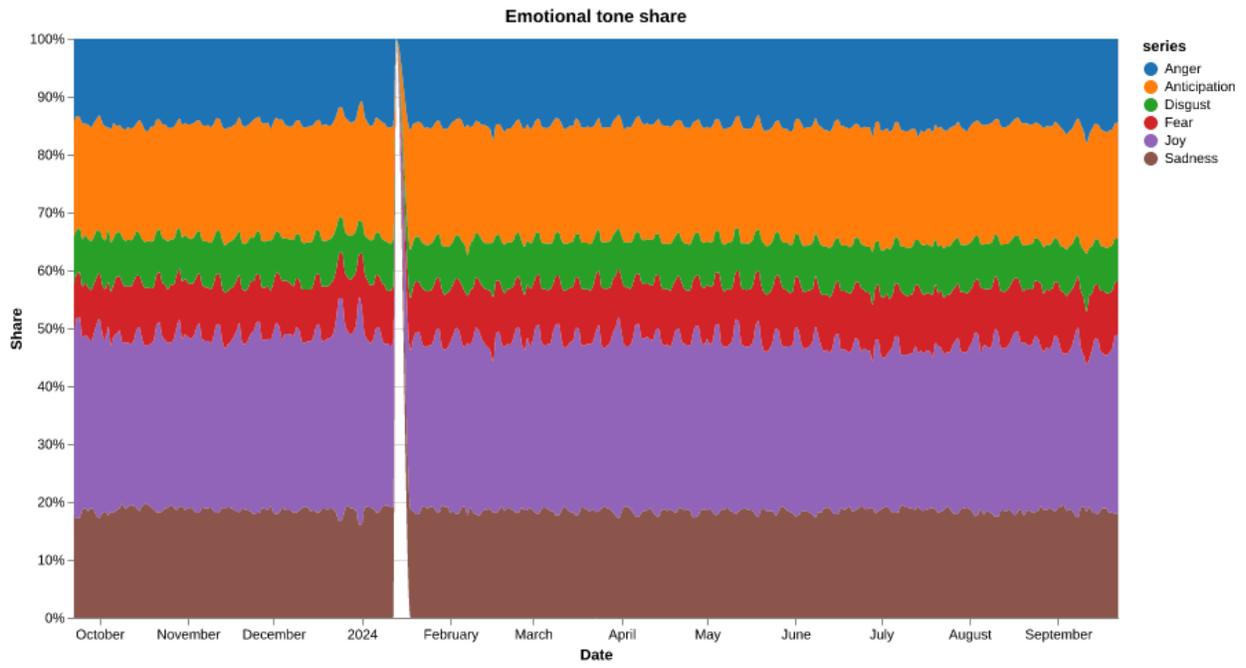

(C)

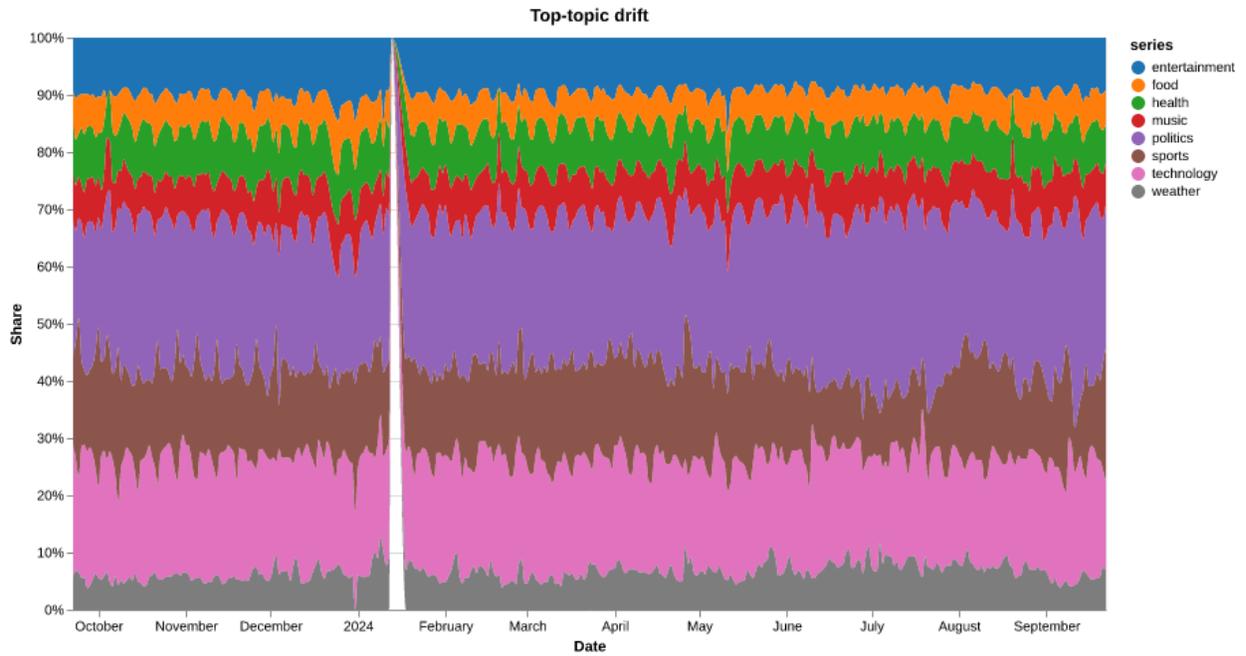

*Figure 1.* LLM-based analysis of over 90 million Mastodon posts for public health surveillance signals

*(September 2023 - October 2024). We collected public Mastodon posts and processed them using GPT-4o mini with a structured schema extracting sentiment, emotional tone, topics, and health-related content. (A) Health signal density showing the fraction of posts classified as health-related (solid line), with sub-traces for posts describing personal health experiences versus public health policy discussions. (B) Emotional tone composition over time, showing the normalized prevalence of Plutchik's core emotions (Joy, Fear, Anticipation, Anger, Sadness, Trust) to capture when discourse shifts toward optimism, anxiety, or frustration. (C) Topic distribution showing the prevalence of major discussion topics (politics, technology, health, etc.) across the study period. Missing data points reflect data collection interruptions.*

Despite their promise, Mastodon and Bluesky have their own share of challenges. The very factors that make them appealing (smaller, community-driven networks) also introduce limitations. Scale is the most obvious issue. Twitter still has hundreds of millions of active users worldwide. Mastodon and Bluesky have far fewer. A Reuters Institute report found that rival networks like Threads, Bluesky, and Mastodon have a reach of 2% or less for news globally [22]. This limited reach complicates digital public health monitoring, as conversations may be rich but might not capture a broad enough population segment to serve as a proxy for population-level trends. Demographic and ideological representativeness presents another challenge. User migration patterns suggest they self-select in ways that make each platform somewhat internally homogeneous. Preliminary evidence indicates that people leaving Twitter chose alternate platforms partly along ideological lines [23,24]. Mastodon's early adopters included many open-source enthusiasts and academics who often skewed left-leaning. Bluesky's community has also been noted for its historically left-leaning culture. Meanwhile, many right-leaning users remained on Twitter or gravitated to different platforms entirely.

Recent work has documented this fragmentation empirically. Comparative analyses of political discourse across platforms show that each discusses similar news events through ideologically distinct lenses [23]. Researchers have begun warning of "echo platforms", i.e. whole social networks that become ideologically homogeneous niches as users migrate to those fitting their preferences [24]. The result is that no single platform today provides the broad public pulse that Twitter once did. Each network offers only a window into a particular population slice. This ideological clustering could affect certain public health signals. A vaccine misinformation outbreak might rage on one platform but be absent on another. Researchers focusing only on one platform might see rational discourse on vaccines while missing vehement anti-vaccine movements propagating elsewhere. For comprehensive public health monitoring, looking at multiple platforms at the same time would be critically important, yet increasingly difficult.

The field of digital public health monitoring must therefore adapt. Researchers may concentrate on use cases that remain viable with smaller, open platforms. Monitoring common, widespread health conditions such as seasonal influenza, COVID-19 trends, or other infectious disease outbreaks, is still plausible because even a small user base can generate useful signals if the disease is prevalent enough. It's important to note that the early digital public health monitoring work using Twitter data referenced above occurred at a time when Twitter wasn't much larger than today's new social media platforms. Disease monitoring doesn't always require a perfectly representative population sample; it needs a consistent proxy indicator of disease activity. Even a niche network can offer valid trend data if the portion of users discussing symptoms fluctuates in parallel with real-world incidence. Past research has shown that relative changes in social media chatter about illness can closely mirror official disease rates [2,3,25].

The fragmentation of users means comprehensive pictures will require integrating multiple data streams. Future digital public health efforts will likely collect data from many channels such as social networks, search engine trends, or online forums. The EU's Digital Services Act may assist here by obligating large platforms to provide data access to researchers [26]. Digital epidemiology could shift from Twitter-centric to a many-platform mosaic approach. With more reliance on decentralized communities, researchers must engage closely to maintain trust and ethical standards. We may also see models of data donation or participatory surveillance take off, where users are voluntarily participating in surveys or tagging posts for research purposes [19,20].

# Funding acknowledgment

MS and SPM received funding from the European Union's Horizon 2020 research and innovation programme – project "Versatile emerging infectious disease observatory – forecasting, nowcasting and tracking in a changing world (VEO)" (No. 874735).12